\documentclass[iop,11pt]{emulateapj}
\usepackage{amsmath}
\usepackage{natbib}

\catcode`_=\active
\newcommand_[1]{\ensuremath{\sb{\mathrm{#1}}}}
\catcode`^=\active
\newcommand^[1]{\ensuremath{\sp{\mathrm{#1}}}}

\newcommand{\cross}{\times}

\newcommand{\curl}{{\bf curl\,}}

\renewcommand{\vec}[1]{{\bf #1}}

\newcommand{\be}[1]{\begin{equation} \label{eq:#1}}
\newcommand{\ee}{\end{equation}}
\newcommand{\ba}[1]{\begin{eqnarray} \label{eq:#1}}
\newcommand{\ea}{\end{eqnarray}}

\newcommand{\pref}{\protect\ref}

\newcommand{\solrad}{\ifmmode{R}_{\rm S}\else${R}_{\rm S}$\fi}
\newcommand{\solmas}{\ifmmode{M}_{\rm S}\else${M}_{\rm S}$\fi}

\newcommand{\tintu}{\ifmmode{\rm erg~cm^{-2}~s^{-1}sr^{-1}}\else 
  erg~cm$^{-2}$~s$^{-1}$~sr$^{-1}$\fi}
\newcommand{\intu}{\ifmmode{\rm erg~cm^{-2}~s^{-1}sr^{-1}\AA^{-1}}\else 
  erg~cm$^{-2}$~s$^{-1}$~sr$^{-1}$\AA$^{-1}$\fi}
\newcommand{\pintu}{\ifmmode{\rm photons~cm^{-2}~s^{-1}sr^{-1}\AA^{-1}}\else 
  photons~cm$^{-2}$~s$^{-1}$~sr$^{-1}\AA^{-1}$\fi}
\newcommand{\phintu}{\ifmmode{\rm photons~px^{-1}~s^{-1}}\else 
  photons~px$^{-1}$~s$^{-1}$\fi}
\newcommand{\fluxu}{\ifmmode{\rm erg~cm^{-2}~s^{-1}}\else 
  erg~cm$^{-2}$~s$^{-1}$\fi}
\newcommand{\velu}{$\,$km$\,$s$^{-1}$}

\newcommand{\wave}{\ifmmode{\lambda} \else$\lambda$\fi}

\newcommand\lta { \mathrel {\hbox to 0pt {\lower 3.7pt \hbox{$\sim$}
      \hss} \raise 1.7pt \hbox{$<$}}}
\newcommand\gta { \mathrel {\hbox to 0pt {\lower 3.7pt \hbox{$\sim$}
      \hss} \raise 1.7pt \hbox{$>$}}}

\newcommand{\ra}{\rightarrow}
\renewcommand{\ra}{\ldots}
\renewcommand{\ra}{..}

\newcommand\tableone{
\begin{table*}
\label{tab:regimes}
\begin{center}
 \caption{Regimes of spectral line polarization in solar plasmas}
 \begin{tabular}{lllllllll}     
 \hline
 \hline
& Ion & Multi- & $|\vec{B}|$ & $\omega_L$ &  $\omega_{Dopp}$ &  $2\pi A$   & $\epsilon_Z$ & $\epsilon_H$ \\ 
& charge & pole & G       & [rad/s]    & [rad/s]       & [rad/s]  & $=\omega_L/\omega_{Dopp}$  & $=\omega_L/2\pi A$ \\
\hline\\
photosphere &0,1&E1 & $10\ra3(3)$ & $1(8)\ra3({10})$ &  $4({10})$ & $5(8)$ & $2({-3})\ra1$ & $0.2\ra60$ \\
chromosphere &0,1&E1 & $10\ra1(3)$ & $1(8)\ra1({10})$ & $4({10})$ & $5(8)$ & $2({-3})\ra0.4$ & $0.2\ra20$\\
corona &$7\ra14$&M1 & $1\ra100$ & $1(7)\ra1(9)$ & $4({11}) $ & 50 & $2({-5})\ra2({-2})$ & $2(5)\ra2(7)$ \\
prominence/filament &0,1&E1 & $5\ra50$ & $7(7)\ra 7(8)$ & $4({10})$ & $5(8)$ & $2({-3})\ra 2({-2})$ & $ 0.14\ra 1.4$ \\
\\
\hline
\end{tabular}
\end{center}
{The table shows data for the photosphere, chromosphere, corona and prominences.  
The lower field strength regions apply to quiet regions, the higher values to 
the strongest concentrations (the darkest regions of sun spots).   Typical
values are listed for the Larmor frequency of atoms and ions ($\omega_L$), the Doppler width  ($\omega_{Dopp}$)
and natural width ($2\pi A$) of the lines in angular 
 frequency units, and the ratios of these parameters.  A reference wavelength of $\lambda=$5000 \AA{} was adopted in making
this table, the values of $\omega_{Dopp}$ vary as $\lambda^{-1}$.
The notation $X(Y)$ means $X\times 10^{Y}$. 
When $\epsilon_Z \ll 1$ the Zeeman intensity profiles are unsplit, broadened, and the 
induced polarization is small.  When $0.1 < \epsilon_H < 100$ the Hanle effect
is important. }
\end{table*}
}

\newcommand\tabletwo{
\begin{table}
\protect\label{tab:pairs}
\begin{center}
 \caption{Close pairs of lines where $``Q,U=0"$ and \ion{Fe}{1} $``V=0"$}
 \begin{tabular}{llllr}     
 \hline
 \hline
\multicolumn{2}{c}{Zero QU} & \multicolumn{2}{c}{Zero V} & $\Delta\lambda$ nm\\ 
Ion       & $\lambda$ nm   & Ion & $\lambda$ nm & \\
\hline
    Mn I &   425.770 & Fe I &   425.6199       &     0.15 \\
    Fe I &    426.53 & Fe I &   425.6199            & 0.91 \\
    Fe I &   730.06 & Fe I &   728.1564       &     1.91 \\
\hline
\end{tabular}
\end{center}
All $V=0$ and $Q,U \ne 0$ lines of \ion{Fe}{1} discussed in the text
are grouped with nearby lines in Tables~1 and 2 of
\citet{Vela-Villahoz+others1994}. Further work is in progress to find lines of  ions 
with ``$V=0$'', other than for \ion{Fe}{1}. 
\end{table}}

\newcommand{\philemail}{judge@ucar.edu}

\newcommand{\figone}{
\begin{figure}[h] 
\epsscale{1.24}
\plotone{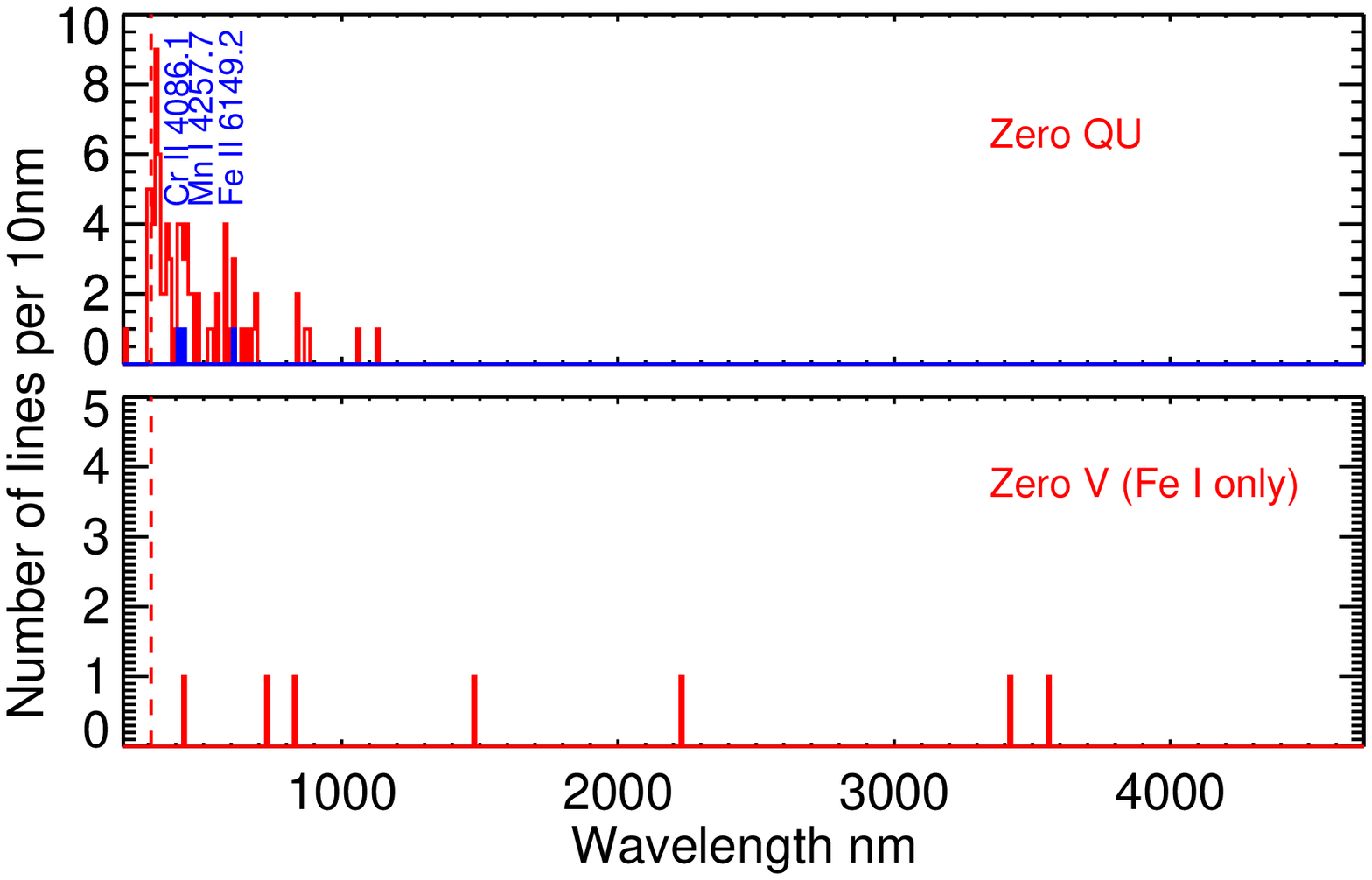}  
\caption{\label{fig:lines} 
The numbers of lines known to have zero linear polarization are shown in the top panel as a function of wavelength, taken from SAVV93.  In red are shown all the transitions listed, and in blue those transitions that are listed as strong enough to observe, and with no known blend.  The lower panel shows transitions with no circular polarization, only of \ion{Fe}{1}, taken from the NIST online database of atomic spectra.  Blue lines shows those for which oscillator strengths are listed by NIST, the red histogram includes all possible transitions.  LS coupling is assumed to be valid. The vertical dashed line shows the atmospheric cutoff at 310 nm.}
\end{figure}
}
\newcommand{\figseven}{
\begin{figure}[h] 
\epsscale{1.18}
\plotone{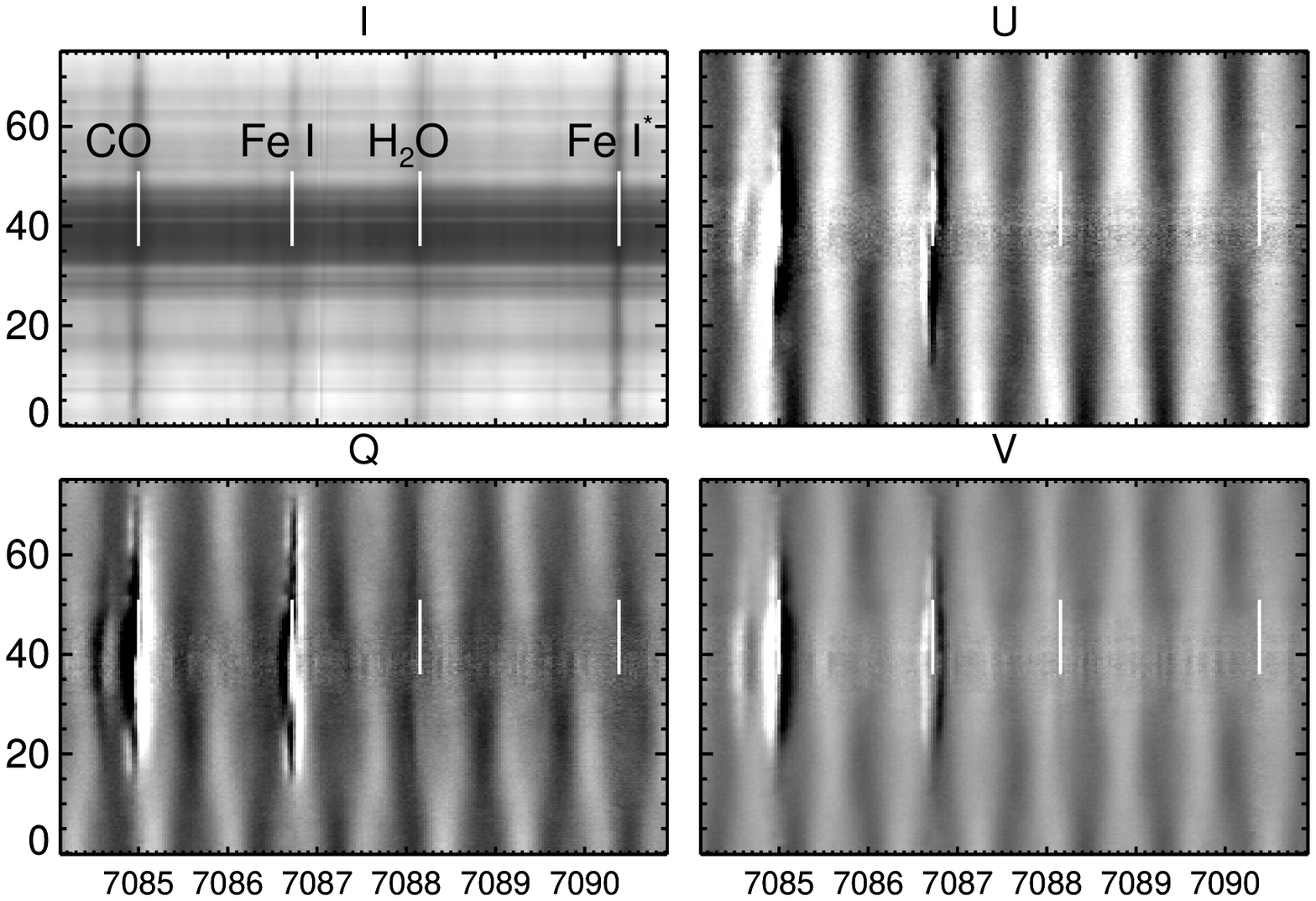}  
\caption{\label{fig:7090} 
Minimally processed Stokes spectra are shown for the 709.0 nm region, obtained 21st August 2014 at the Dunn Solar Telescope using the SPINOR spectro-polarimeter. Standard calibration procedures were applied with no attempts to "massage" these data. The abscissa is wavelength in \AA, the ordinate position along the spectrograph slit (arc seconds). The middle dark portion is the umbra (darkest region) of a small sunspot in NOAA active region 12147.  The 709.04 nm transition of \ion{Fe}{1} (${\rm 5s\,^5F_1  \rightarrow  4p\, ^5D^O_0}$, both with the ${\rm 3p^6 3d^7 (^4F)}$ core), shows zero polarization at the sensitivity levels achieved in these observations. The $Q,U$ panels are color scaled between $\pm 0.02I$, $V$ is shown between $\pm 0.1I$.}
\end{figure}
}

\newcommand{\figten}{
\begin{figure}[h] 
\epsscale{1.18}
\plotone{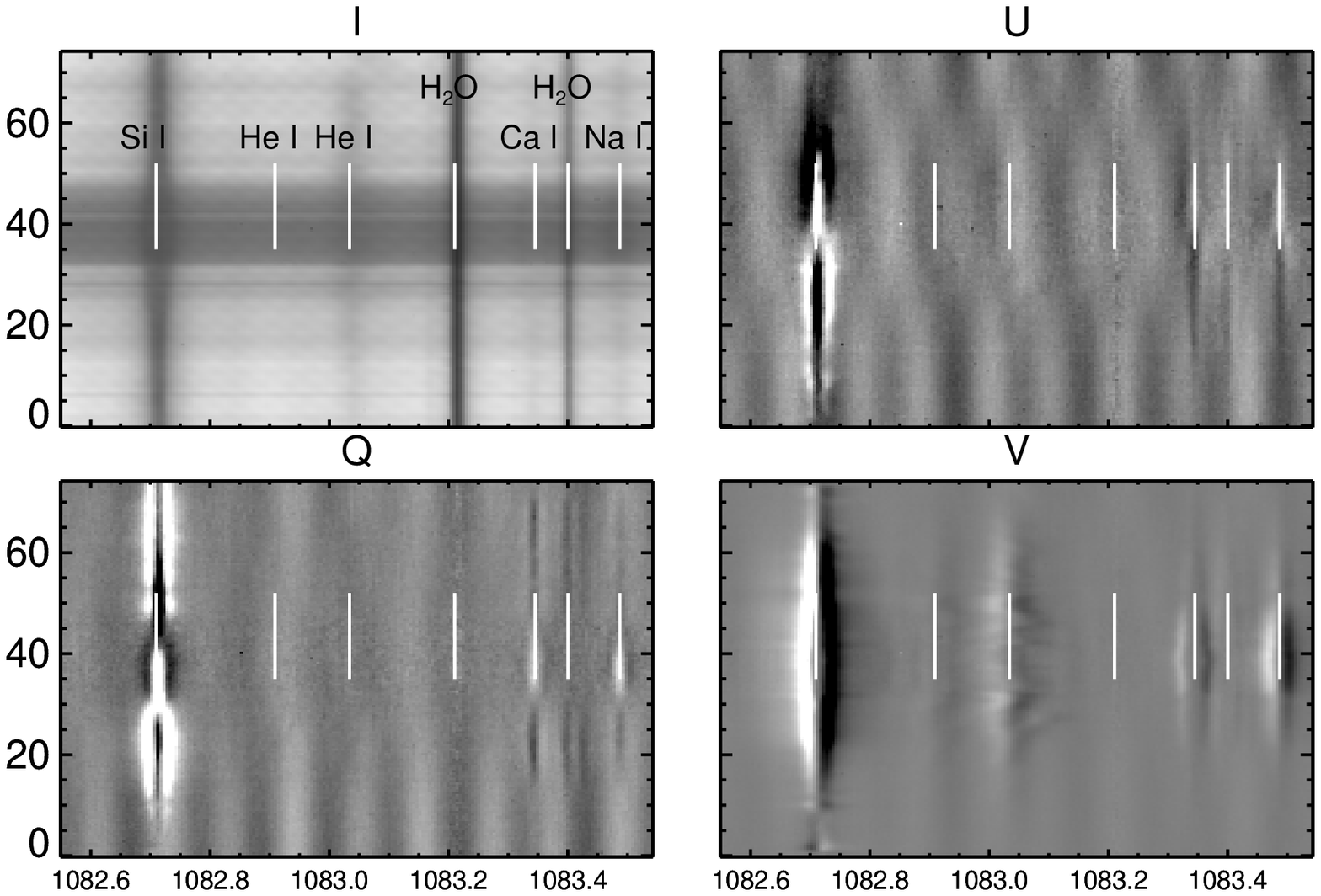}  
\caption{\label{fig:10830} 
Simultaneous SPINOR spectra  as in \pref{fig:7090} for the region near the 1083 nm multiplet of \ion{He}{1}.  All solar lines show polarization signals, there is also a signal from the telluric (unpolarized) line of H$_2$O (see the panel for Stokes $Q,U$ only near 1083.4 nm and $y=$) which must be due to an inaccurate polarization calibration.  Notice that, unlike the 709 nm region shown in Figure~\pref{fig:7090}, the $QUV$ profiles (e.g., of \ion{Si}{1} 1082.7 nm) are consistent with solar Zeeman patterns, the polarization calibration is easier at infrared wavelengths.}
\end{figure}
}

\shortauthors{Judge}
\shorttitle{Atomic physics and solar polarimetry}

\slugcomment{}

\begin{document}

%
%

\title{Atomic physics and modern solar spectro-polarimetry}
\author{  
Philip G. Judge}
\affil{High Altitude Observatory,\\
       National Center for Atmospheric Research\thanks{The National %
       Center for Atmospheric Research is sponsored by the %
       National Science Foundation},\\
       P.O.~Box 3000, Boulder CO~80307-3000, USA; \philemail}

%
%

\begin{abstract}

Observational solar physics is entering a new era with the advent of
new 1.5 m class telescopes with adaptive optics, as well as the Daniel
K. Inouye 4 m telescope which will become operational in 2019. Major
outstanding problems in solar physics all relate to the solar magnetic
field.  Spectropolarimetry offers the best, and sometimes only, method
for accurate measurements of the magnetic field.  In this paper we
highlight how certain atomic transitions can help us provide both
calibration data, as well as diagnostic information on solar magnetic
fields, in the presence of residual image distortions through the
atmosphere close to, but not at the diffraction limits of large and polarizing 
telescopes.  Particularly useful are spectral lines of neutrals and
singly charged ions of iron and other complex atoms.  As a 
proof-of-concept, we explore atomic transitions that might be used to
study magnetic fields without the need for an explicit calibration
sequence, offering practical solutions to the difficult challenges of
calibrating the next generation of solar spectropolarimetric
telescopes.  Suggestions for additional work on atomic theory and
measurements, particularly at infrared wavelengths, are given.  There is some promise for continued symbiotic advances between solar physics and atomic physics.
\end{abstract}


\section{Introduction}

The fields of solar  and atomic physics have enjoyed decades of
fruitful collaborations \citep[e.g.][]{Gabriel+Jordan1971,Dufton+Kingston1981}.  The Sun is our
best ``laboratory'' for studying the behavior of an archetypal,
nearly-ideal plasma under conditions of very high magnetic Reynolds
numbers \citep[e.g.][]{Parker1979}. The Sun also spans wide ranges of plasma $\beta = 8\pi P/B^2$.  On average, 
$\beta \gg 1$ in the interior, $\beta \approx 1$ in the atmosphere (from which the bulk of the solar radiation
escapes), and $\ll 1$ in the solar corona.  

The Sun exhibits  {\em complex} behavior, i.e.  
patterns emerge from non-linear governing equations of
motion, a result that appears larger than the sum of the parts.
Yet, seven decades after the development of 
magneto-hydro-dynamics, the simplest model capable of entertaining
such behavior, we still are unable to answer the deceptively simple
question: {\em How does the Sun regulate its strikingly ordered and
  ever-changing magnetic field?}.

Solar differential rotation, in spite of (or perhaps because of) 
turbulent convection beneath the solar surface,
leads to well-known 
patterns of magnetic structure, such as the remarkable ``sunspot cycle''.
Every eleven years the entire global solar magnetic field reverses.  This occurs in
a system in which the global magnetic diffusion time is some $10^9$
years.  We also do not know the physical reasons why the Sun is obliged to
form {\em spots}.  These intense
concentrations of magnetic field, too often taken for granted, were first studied in Galileo's era.
But why does magnetic flux appear in such intense
concentrations in the Sun, 
sometimes exceeding field strengths in equipartition
with the convection?

\tableone

Such  are the nature of some of the major unsolved problems in solar physics. 

\section{The continuing need for observations}

Following decades of exponential advances in computations, it might be
surprising that numerical experiments are far from providing us with
an {\em ab-initio} understanding of the Sun's behavior. However, this is because of
the extreme range of scales involved.  Consider the governing equation
for the magnetic field in magneto-hydrodynamics, readily derived from
Faraday's Law of Electromagnetic Induction and Ohm's Law (kinetic
collisional dynamics):
\begin{equation} \label{jeqinduction}
{ {\partial \vec{B}}\over {\partial t}} = \curl (\vec{u} \times \vec{B}) +
\eta \nabla^2 \vec{B}, 
\end{equation}
where, in the solar interior for example, the magnetic diffusivity $\eta$ is $\approx $ $10^4$
cm$^2$s$^{-1}$.  Over scales of a fraction of a solar radius $\ell
\approx 10^{10}$ cm, ordered flows are $\approx 0.1$ \velu{}.  Thus the
``magnetic Reynolds number'' is $|\vec{u}| \ell / \eta \approx
10^{10}$.  3D numerical simulations typically have $R_M \lta 10^4$.
{\em The numerical range of scales is some 6 orders of magnitude
  smaller than the physical range.  } The Lorentz force, $(\curl
\vec{B}) \cross \vec{B}$, increases with inverse scale length, 
not allowing us to invoke a ``simple'' turbulent  
cascade \citep{Parker2009}.  Therefore, the fundamental
physics of magnetic regeneration -- the ``dynamo'' problem -- implies that 
\begin{quote}
{\em solar physics remains an observationally-driven science.}
\end{quote}
To measure magnetic fields, {\em spectropolarimetry},
developed from the 1960s, is the most powerful tool at hand.
With the advent of new 1.5-meter to 4-meter class 
telescopes, spectropolarimetry is poised to make important
breakthroughs. These new telescope systems off 
higher angular resolution, larger photon flux, access to thermal infrared regions and 
coronagraphic capabilities.  With modern adaptive optics systems \citep{Rimmele+Marino2011}, 
these 
instruments will permit us to study evolving 
surface magnetic fields across the physical scales of interest.  

\figseven

Observations and 
and numerical experiments yield ``effective'' (i.e. non-kinetic, or ``turbulent'') 
diffusivities of $\approx 10^{12}$ cm$^2$s$^{-1}$
\citep[e.g.][]{Berger+others1998,Cameron+Voegler+Schuessler2011}, sufficient to account for the 11 year evolution time of the solar magnetic field.  But these diffusivities as yet have no solid justification in physics \citep{Parker2009}.  But if we accept these values, 
with convective speeds $u$ of order 3 \velu{}, this diffusion coefficient 
$\eta \approx  \frac{1}{3} u \ell$ implies $\ell \approx 100 $ km. 
The 4-meter Daniel K. Inouye Solar Telescope (DKIST, previously ATST,
\citealp{Keil+Rimmele+Wagner2009}) will 
resolve scales down to $\approx 20$ km.   New DKIST observations will therefore 
help us answer the most pressing questions regarding the evolution of 
solar magnetism.  

\figten

Before proceeding, we display some data in figures~\pref{fig:7090} and
\pref{fig:10830}. They show Stokes (polarized) spectra.
In natural sources like the Sun, polarized light occurs through 
averages of incoherent packets of photons.  
In solar work it is therefore traditional
to use the Stokes parameters $(I,Q,U,V)^T$ written as an array ${\bf
  S}$, which can be simply related to the coherency matrix. Here
 $I$ is the intensity, $Q,U$ are
the linear polarization parameters, $V$ circular. 
It is possible to give an operational definition of 
${\bf S}$ in terms of an ideal linear polarizer and
retarder set at different angles relative to a fixed
direction in the plane of polarization
 \citep[see, e.g.,][ch.~1]{Landi+Landolfi2004}. 

The data shown were taken with the SPINOR
instrument at the Dunn Solar Telescope (DST), which in these data has
an angular resolution of $\approx 400$ km at the solar surface.  
The DST is fed by a
plane heliostat and the optical system is far from symmetric (with consequences discussed more below). These data were obtained
simultaneously under typical observing conditions at red and infrared
wavelengths using a state-of-the-art adaptive optics system.  They
show several properties: real solar signals, ``cross-talk" from $V$ to
$Q$ and $U$ (the 709 nm region has lines with non-physical
antisymmetric $Q,U$ profiles), optical fringing, at the level of 1\%,
and some unphysical polarization in telluric (H$_2$O) lines.  The non-solar signals are the main concern of this article.  Several ways in which the modern telescopes in fact make polarimetry more challenging are discussed below. 

\section{New regimes for spectro-polarimetry}

Remarkably, the Sun is simply {\em not bright enough} to tackle the
demands of measurements of solar magnetism \citep{Landi2013}.
Solar fields are weak compared with laboratory fields, and Zeeman
splittings are small compared to Doppler widths (i.e. $\epsilon_Z \ll
1$ in Table~\pref{tab:regimes}). Information on the magnetic field is
therefore encoded mostly through {\em spectral line polarization}, 
the Zeeman splitting in intensity profiles 
being far smaller than the line widths.

The Sun's visible ($\lambda \approx$ 5000 \AA) intensity is $I_\lambda
\approx B_\lambda(T$=5500K) \intu, where $B_\lambda$ is the Planck
function.  To compute the photon flux density from a solar area
subtending a solid angle of $\varpi$ steradians, we have $f_\lambda=\varpi
B_\lambda (\lambda/hc) $ photons~cm$^{-2}$\AA{}$^{-1}$s$^{-1}$.  For a
telescope of diameter $D$ cm, the flux from this area, 
integrated over the aperture is
\be{flux} F_\lambda = \varpi B_\lambda \frac{\lambda}{hc} \frac{\pi
  D^2}{4} {\rm photons~\AA{}^{-1}~s^{-1}}.  
\ee 
If we critically 
sample the solar image spatially {\em at the diffraction limit}, i.e.
at half of the diameter of the telescope point spread function, then 
$\varpi = \left (1.22\lambda/2D\right)^2$, and 
\be{fluxdl}
F_\lambda^{DL} = \frac{\pi }{4} B_\lambda \frac{\lambda}{hc} \frac
{(1.22 \lambda)^2}{4} = 3.8\times10^8 \ {\rm photons~\AA{}^{-1}~s^{-1}},
\ee 
independent of the telescope aperture.  Photospheric
Doppler widths are 2-3 \velu{}.  Spectrographs with resolutions $R
\gta 200,000$ ($\equiv 1.5 $ \velu, $\equiv 0.024$ \AA{} @ 5000 \AA) are
typically used. But Zeeman-induced polarization
is of order $ \epsilon_Z {dI}/{d\omega} \equiv \epsilon_Z I'$ and $ \epsilon_Z^2
{d^2\!I}/{d\omega^2}$ respectively, in the limit $\epsilon_Z \ll
1$ \citep[e.g.][]{Casini+Landi2008}.  Polarimetry requires slightly 
higher spectral 
resolution than intensity spectroscopy, the profiles being 
(to lowest order) wavelength 
derivatives of the intensity profile.  Let us use a
pixel width $ \lta \lambda/ 2R$ or
$\lta 12.5$ m\AA{}.  With a total system efficiency of ${\cal E}$, the
flux per 12.5 m\AA{} pixel is 
\be{fluxdlp} F_p^{DL} = 4.7\times10^6 {\cal
  E} \ {\rm photons~px^{-1}~s^{-1}} 
\ee
If ${\cal E}=0.05$, $F_p^{DL}\approx 2.4\times10^5$ {\rm
  photons~px$^{-1}$~s$^{-1}$}, and a photon counting signal-to-noise
ratio (SNR) of $\approx 480 \sqrt{t}$, where $t$ is the integration time in
seconds.  To complete these SNR estimates, we must consider additionally:
\begin{itemize}
\item The Sun's atmosphere itself changes during
  integrations.  For a 4-meter aperture, the angular resolution is
  $A 1.22 \lambda/D\equiv 20$ km at the Sun's surface, where
$A=1.496\times10^{13}$ cm is one A.U. 
  Using the sound speed $c_S\approx 7$ \velu, the integration times 
  are limited to $1.22 \lambda A /Dc_S \approx 3$ s, varying inversely with telescope aperture $D$. 
\item At least four measurements must be made during the integration times
to recover the four components $S_i=I,Q,U$ and $V$ of $S$. Hence
integration times must additionally be $\lta 1.22 \lambda A /4Dc_S \approx 0.8$ s.
\item We can only measure 
linear combinations of intensity with $Q,U$ and $V$ (see equation
\pref{eq:modstate} below).
Typically, since 
  $Q,U \propto \epsilon_Z^2$ or $V \propto \epsilon_Z$, then 
$|Q,U,V| \ll  I$. The SNRs of $Q,U$ and $V$ are therefore 
  factors $\epsilon_Z^2$ and $\epsilon_Z$ smaller 
for $Q$ (and $U$) and $V$ than for $I$, respectively. 
\item Entire line profiles are used to
infer magnetic field parameters, using 
ten or more Doppler widths of spectrum, so that  
$n_\lambda \approx 20$ wavelength pixels.
\end{itemize}

Therefore the SNRs at 5000 \AA{} are
\be{snr}
{\rm{SNR}} \approx 2000 {\frac{400}{D}}\epsilon_Z^\alpha, \ \ 
{\rm \alpha =0 \ (I), 2 \ (Q,U),\ 1 \ (V)},
\ee
varying with wavelength as $\lambda^3 $ at visible and infrared
wavelengths, using the Rayleigh-Jeans limit of the
Planck function ($B_\lambda \propto \lambda^{-1}$), and noting that
$\epsilon_Z \propto \lambda$ (Table 1).  To study evolving fields of 10 G,
characteristic of the quiet Sun, $\epsilon_Z \approx 10^{-3}$
(Table~\pref{tab:regimes}).  In this case we find SNRs of just 2 and
$2\times10^{-3}$ for circular and linear polarization respectively,
for $\lambda=5000$\AA.  From this simple analysis, we can conclude the
following.

\begin{enumerate}
\item Solar Zeeman spectropolarimetry should be done far from the
  diffraction limit, at the longest
  wavelengths observable yet compatible with the desired angular resolution. 
\item Very accurate calibrations of instrumental polarizations are
  needed.
\end{enumerate}

Both atomic- and astro- physics limit the available transitions we can
use for item 1.  For example, the visible solar spectrum is dominated
by lines of \ion{Fe}{1}.  The instruments we can develop limit our
choices in item 2. GREGOR is an on-axis 1.5-meter telescope with very
small instrument polarization \citep{Denker+others2012}.  The
1.6-meter New Solar Telescope \citep{Goode+Cao2012} and 4-meter DKIST
are off-axis designs with considerable telescope polarization.  The NST and DKIST 
unobstructed off-axis designs are favored 
for low scattered
light, but they come at a cost.  Incoming polarized light, distorted
by differential refraction in the Earth's turbulent atmosphere
(``seeing''), is mixed before reaching the polarization analyzer.
Under these conditions spurious polarization signals are determined by
the statistics of the seeing, setting lower limits on the sensitivity
of the measurements.

Fortunately, atomic physics can help with these difficulties, by
providing atomic transitions for which the solar polarization
properties are known, no matter the state of the emitting plasmas.  Henceforth, we
will assume pure LS coupling unless specified otherwise.

\section{Solar polarimetry in a nutshell}

In adopting the Stokes description, the measurement
process can be written as matrix products, each representing
an element in the optical system
\citep[e.g.][]{Seagraves+Elmore1994}.  
The goal is to recover the solar
{\bf S} entering a telescope.   Each optical element can be represented by a
matrix. A  $4\times4$ ``M\"uller" matrix is used to characterize the change in 
{\bf S} for each optical element, but the mathematics can also include larger
matrices as needed to 
handle beam-splitters and different modulation schemes 
\citep{Seagraves+Elmore1994}.  

When stripped to the bare bones, the essence of the polarization
measurement process can be written as follows. The incoming solar 
light is modified by the telescope
and optical feed system ({\bf X}) and 
passed through an optical modulator (e.g., a rotating retarder) 
which alters systematically and
repeatedly the polarization state of the light.  An analyzer element 
(linear polarizer) 
in front of the detector converts the modulated polarized light into
an intensity.  
The combined modulator-analyzer and other elements (e.g., spectrograph) can be conceptually written by a $4\times N$ matrix
{\bf M}.  This matrix 
produces a $1\times N$ ($N\ge 4$) vector {\bf C} of counts
on a detector:
\be{algebra}
{ \bf  C =  (M X) S}
\ee
Finally, {\bf S} is recovered from
\be{algebrai}
{ \bf S =  (M^T M X)^{-1}  M^T C }
\ee
One critical property of {\bf C} is not evident from this algebra,
namely that $Q,U,V$ always occurs in linear combination with $I$, since 
within 
a gain and dark correction, 
\be{modstate}
C_j=I+a_jQ+b_jU+c_jV
\ee
with $a_j,b_j,c_j$ constant.  Therefore, as summarized above, 
noise in {\bf C} is dominated
by noise in $I_j$ which is $ \approx I$ when $\epsilon_Z < 1$.

Solar physicists would be very happy with this situation!  In the
imperfect real world, we face serious additional challenges:
\begin{enumerate}
\item {\bf S} suffers from high frequency distortions as solar light
  passes through Earth's atmospheric turbulence. At any given time
  {\bf S = S_\odot + $\delta$S}, but only statistical properties of
  $\delta${\bf{}S} can be determined.
\item Modulation is done in time, the states $j$ in
  equation~(\pref{eq:modstate}) each experience different realizations of
  $\delta${\bf S}.
\item There will be residual errors in the telescope matrix {\bf X}
  and the remaining matrices {\bf M}.
\item The detector counts {\bf C} in
equation~(\pref{eq:modstate}) will have dark, gain residuals and
  other imperfections.
\end{enumerate}
The problems faced can be illustrated using departures from the simplest case ${\bf  X=\mathbf{1}}$.
We seek accurate measurements of the solar input Stokes vector ${\bf S
  \approx (I,\epsilon_Z^2I^{\prime\prime},
  \epsilon_Z^2I^{\prime\prime}, \epsilon_ZI^{\prime})}^T$.  The effect
of ${\bf X} \ne {\mathbf 1}$ is to ``mix'' the $I,Q,U,V$ components
before entering the modulator and downstream optical elements. Some
residual mixing of this type is seen particularly in Figure 1, where $Q$ and $U$ clearly have the character of $V\propto I'$ and not $I''$.  We now
examine how atomic physics can help side-step some of this mixing.

\section{How atomic physics can help polarimetry in solar physics}

\subsection{Measurement of longitudinal fields only}

Suppose that we want to measure not the full Stokes
vector ${\bf S}$, but just the Stokes components $I$ and $V$.  This is the essential 
idea behind the original ``longitudinal 
magnetograph", motivated by the fact that the $V$ signal 
is first order in the small quantity $\epsilon_Z$,  allowing us to measure 
the line-of-sight components of the solar magnetic field 
\citep[e.g][]{Babcock+Babcock1952,Babcock1953}.   If ${\bf X} = \mathbf{1}$ then there 
s no issue, any spectral line which has a non-zero Land\'e g-factor can be used.
However, if we are using a polarizing telescope ${\bf X} \ne \mathbf{1}$,  then 
equation~(\pref{eq:modstate}) implies that, to recover $I$ and $V$, we must know 
all components of the matrix ${\bf X}$.

\figone

\citet[][henceforth SAVV93]{Sanchez-Almeida+Vela-Villahoz1993}
proposed a solution to this problem without full knowledge of ${\bf
  X}$, prompted in part by a study 
of polarization properties of 
the \ion{Fe}{2} line at 614.92 nm
\citet{Lites1993}. When the {\bf X} matrix satisfies come commonly encountered symmetry properties,  measurements of continuum and lines known {\em
  a priori} to generate zero linear polarization, the needed elements
of ${\bf X}$ can be algebraically eliminated to a high level of
accuracy (see eqs. 4 and 7 of SAVV93).
The particular transitions of interest are characterized by peculiar Zeeman patterns where a compensation occurs between $\sigma$ and $\pi$ components, causing 
the transfer equations for the Stokes parameters $Q$, $U$ to be decoupled from those for $I$ and $V$. When the boundary values for $Q$ and $U$ are also zero (deep in the atmosphere) the emergent linear polarization is then zero.  These transitions are:
\begin{eqnarray}
^4{\rm D}_{1/2} &\rightarrow&\  ^{2S+1}{\rm L}_{1/2}, \nonumber \\
^6{\rm G}_{3/2} &\rightarrow&\  ^{2S+1}{\rm L}_{1/2}, \nonumber \\
^{22}{\rm O}_{3/2} &\rightarrow&\  ^{2S+1}{\rm L}_{1/2},\label{eq:noqu}\\
&\ & \hskip -26pt {\rm where \ } ^{2S+1}{\rm L \ne  ^4\!D} \nonumber
\end{eqnarray}
(See also table 9.4 of \citealp{Landi+Landolfi2004}). 
The latter condition ensures that the LS-coupled Land\'e g-factors are
non-zero, and therefore $V\ne 0$.  
A list of these lines, assuming LS coupling is valid, is given in Table 1 of 
\citet{Vela-Villahoz+others1994}. 
The transitions belong only to atomic systems with odd
numbers ($n=3, 5, 7...$) of electrons, thus excluding the rich
spectrum of \ion{Fe}{1} from the Sun's photosphere.

Of 86  such lines listed by \citet{Vela-Villahoz+others1994}, just 3 are unblended, lying above
the Earth's atmospheric cutoff at 310 nm and which belong to an abundant
element ($> 10^{-6}\times$ hydrogen).  Figure~\pref{fig:lines} shows,
in the upper panel, those lines compiled by \citet{Vela-Villahoz+others1994} that satisfy the
constraints of equation~(\pref{eq:noqu}).  The three lines in the
visible region which remain sufficiently unblended to be of real
practical use, are marked in blue.  

\subsection{Lines with linear but no circular polarization?}
\label{subsec:nocirc}

The analysis of SAVV93 suggests that, if lines with $V=0$ but with non-zero 
$Q$ and $U$ genuinely
exist, then their analysis might be extended to try to recover the
full Stokes vector {\bf S}.  But citing \citet[][in particular figure 9]{Makita1986}, SAVV93 
note that even if the Land\'e g-factor is zero, magneto-optical effects can produce circular polarization.  Such polarization is generally small \citep[][section 9.22]{Landi+Landolfi2004}, being of order $\epsilon_Z^4$ for Stokes $V$ in the 
weak field case. 

\citet{Landstreet1969} searched Moore's 1945 revised multiplet table
for LS coupled transitions with zero Zeeman splitting in the presence
of magnetic fields.  The $g=0$ levels, when connected with either a $J=0$ of $g=0$ level generate no Zeeman-induced polarization {\em at all}
since the levels are unsplit.
The 709.04 nm transition of \ion{Fe}{1} (${\rm 5s\,^5F_1 \rightarrow
  4p\, ^5D^o_0}$, both with the ${\rm 3p^6 3d^7 (^4F)}$ core) is an
example, the absence of polarization of this line is seen in Figure~1, showing that such lines can useful as limited checks of calibration procedures. 

However, we need transitions for which $V=0$ but for which $Q,U$ are
non-zero, in order that we can determine the needed 
elements of {\bf X}.    Table 9.4 of \citet{Landi+Landolfi2004} 
lists several such transitions, which are mostly 
spin-forbidden, and most of which also require $\Delta{\rm L}=2$:
\begin{eqnarray}
^6{\rm P}_{3/2} &\rightarrow\  ^{4}{\rm F}_{5/2}, \nonumber &\ (\Delta {\rm S}=1, \Delta{\rm L}=2)\\
^5{\rm D}_{2} &\rightarrow\  ^{3}{\rm G}_{3}, \nonumber &\ (\Delta {\rm S}=1, \Delta{\rm L}=2)\\
^7{\rm D}_{1} &\rightarrow\  ^{5}{\rm F}_{2}, \nonumber &\ (\Delta {\rm S}=1, \Delta{\rm L}=1)\\
^8{\rm D}_{5/2} &\rightarrow\  ^{6}{\rm G}_{7/2}, \nonumber &\ (\Delta {\rm S}=1, \Delta{\rm L}=2)\\
^5{\rm F}_{2} &\rightarrow\  ^{5}{\rm H}_{3},  &\ (\Delta {\rm S}=0, \Delta{\rm L}=2) \label{eq:nov}\\
^7{\rm F}_{2} &\rightarrow\  ^{7}{\rm H}_{3}, \nonumber &\ (\Delta {\rm S}=0, \Delta{\rm L}=2)\\
^7{\rm F}_{3} &\rightarrow\  ^{5}{\rm H}_{4}, \nonumber &\ (\Delta {\rm S}=1, \Delta{\rm L}=2)\\
^8{\rm F}_{3/2} &\rightarrow\  ^{6}{\rm G}_{5/2}, \nonumber&\ (\Delta {\rm S}=1, \Delta{\rm L}=1)
\end{eqnarray}

The conditions for the existence of lines with ``$V=0$'' and ``$Q,U \ne 0$'' are interesting.  In order 
to produce transitions of electric dipole (E1) character 
at all, the level(s) involved must be mixed by spin-orbit
or similar interactions, since under strict LS coupling these are spin- and/or total angular momentum-
forbidden.  But this also means that the Land\'e g-factors of the mixed levels must be non-zero.
(Alternatively, such transitions might be  magnetic dipole, or 
quadrupolar transitions, but these will be much weaker for neutrals or singly ionized ions).  
When such E1-type mixing occurs, then with \citep{Cowan1981}  
\be{mixing}
|\alpha J\rangle = \sum_{\gamma SL}
|\gamma SLJ\rangle
\langle \gamma SLJ|\alpha J \rangle
\ee
the Zeeman splitting of the mixed level is 
\be{mixingg}
g_{\alpha J}= \sum_{\gamma SL}
g_{SLJ}\ 
|\langle \gamma SLJ|\alpha J \rangle|^2.
\ee
For a spin-forbidden (SF) transition, the same mixing induces 
the radiative transition via a fully permitted
E1 transition. For illustration, if just one
level is mixed with one other, 
say the  $|^{5}{\rm F}_{2}\rangle$
level is actually 
$|^{5}{\rm F}_{2}\rangle +\epsilon|^{7}{\rm F}_{2}\rangle$, $|\epsilon| \ll 1$,
then the E1 line strength ${\cal S} (\propto gf)$ is 
\be{mixingf}
{\cal S}(^7{\rm D}_{1} \rightarrow\  ^{5}{\rm F}_{2})  \approx 
\epsilon^2 \ {\cal S}(^7{\rm D}_{1}
\rightarrow\  ^{7}{\rm F}_{2})
\ee
It is clear that there is in principle {\em no transition with finite 
Q,U and zero V}, since the conditions given by (\pref{eq:nov}) and (\pref{eq:mixingf}) requires
a finite mixing coefficient which leads to an, albeit small, modification
of a Land\'e g-factor in equation~(\pref{eq:mixingg}).   
Since the Land\'e g-factor of a transition is the combination of
g-factors of the two atomic levels involved, each case must be 
examined to see the effect of the mixing on the Zeeman patterns.  
But it seems likely,
that transitions might be found which will have {\em small enough} $g$-factors and 
{\em small enough} magneto-optical effects that they have 
{\em very small $V$}, at the same time having a finite $Q,U$.  Equivalent lines for the ``zero $Q,U$'' case are listed in Table 2 of 
\citet{Vela-Villahoz+others1994},

We will assume that $V$ is small
enough to lie within the noise of solar measurements henceforth.  This
assumption will be examined in a later publication. 

To begin exploring such transitions, we examine the spectrum of \ion{Fe}{1}
which dominates (by number) the photospheric spectrum of the Sun.  The
transition $^7{\rm D}_{1} \rightarrow\ ^{5}{\rm F}_{2}$ has no entries
in the NIST atomic database, but there are semi-empirical $gf$-values
from \citet{Kurucz1988}.  Examples of these \ion{Fe}{1} lines in the
solar spectrum include 425.6199 nm (log~$gf=-2.4$), 728.1564
(log~$gf=-4.2$), 830.7606 (log~$gf=-5.5$), although the latter two
lines are blended with telluric H$_2$O.  There are others 
predicted at infrared wavelengths including 1.478302, 2.234095,
3.423527, 3.558674, 6.439891 $\mu$m with log~$gf$ between -3 and -4.  The transitions are marked in Figure 3.

\subsection{Feasibility study for vector polarimetry}
\label{subsec:vecpol}

Here we generalize the approach of SAVV93.  Consider that we
can observe two lines close together in the spectrum, one known to
produce ${\bf S_1 =} (I_1,0,0,V_1)^T$, the other ${\bf S_2 =}
(I_2,Q_2,U_2,\epsilon)^T$, with $\epsilon \ll 1$.  For convenience, we will assume
$\epsilon=0$ below, i.e. it lies below the noise levels. 
Like SAVV93,
we will assume we can measure the neighboring continuum with ${\bf S_0
  =} (I_0,0,0,0)^T$.  Now make measurements of these lines and
continuum wavelengths simultaneously, each one obeying
equation~(\pref{eq:algebra}).  Each of the three arrays ${\bf S_i}$
yields an array of (at least) four counts {\bf C}
\be{grand}
{\bf C_i  =  (MX)_i S_i}.
\ee
In a weakly polarizing telescope ${\bf X}$ obeys certain symmetries 
(see equation A3.a of SAVV93), leaving just 7 independent matrix elements 
(one on the diagonal and 6 off-diagonal). 
 {\bf X} must be of the form
\vskip 8pt
\[
{\bf X}= g \begin{bmatrix}
1 & a & b & c\\
a & 1 & d & e\\
b & -d & 1 & f\\
c & -e & -f & 1\\
\end{bmatrix}
\]
\vskip 8pt
\noindent usually with 
$|a|,|b|,|c|,|d|,|e|,|f| \ll 1$ 
and $g$ can be considered a calibration factor (counts per unit intensity), assumed fixed during the observations.  
The small values of $|a|\ldots |f|$ are not important, but the symmetries are, we will obtain solutions
only when there are at most seven variables $a\ldots g$.
{\bf X} matrices for the DKIST 
have been studied by \cite{Harrington+Sueoka2016}, broken down into primary, secondary and Coud\'e feed 
optics and finally instrument 
optics.   They conclude that after the Gregorian (secondary mirror, ``M2'') reflection  

\begin{quote}
``Only the IQ and QI terms have substantial amplitude [$\approx 0.27\%$] the lack of cancellation from M2 reversing the sign of the reflection.''
\end{quote}
Simulated images across the 5 arcminute FOV show that the matrices are  
close to the weakly polarizing form. Thus, a modulator placed immediately after 
M2 would satisfy the requirements for application of the proposed formalism\footnote{This is not the configuration anticipate during the commissioning phase of DKIST.}.  

Further down the optical chain, before the proposed modulators for the
VTF, ViSP, and DL-NIRSP instruments, the {\bf X} matrices of
\cite{Harrington+Sueoka2016} appear to exhibit the symmetries of the
above weakly polarizing matrix, but $d,e,f$ ($QUV$ to $QUV$ crosstalk
terms) can become large as the telescope rotates while tracking the
Sun.  We will assume henceforth that our model, relying only on the
symmetry properties, can be applied to DKIST.

Thus {\em our goal is to determine all the independent elements ${\bf S_i}$, 
relative to the continuum intensity, plus the six coefficients $a$-$f$ of the 
{\bf X} matrix, given the twelve measurements made with the specific input 
Stokes ${\bf S_0, S_1, S_2}$ with the properties detailed above. }

Let us assume that the counts have been 
dark and gain corrected.
``Analyzing'' (i.e. applying calibrated matrices {\bf M} after the modulator stage; this could be done infrequently in the fixed frame of the Coud\'e lab) 
  without knowledge of {\bf X} produces the set of "measured" Stokes parameters {\bf c'_i} using 
\be{demod}
{\bf c'_i = (M^TM)^{-1} C_i  =  X_i S_i}.
\ee
For the continuum measurements
(${\bf S= S_0}$), dividing all intensities ${\bf c'}$ by the 
measured intensity $ = g i_0$ then 
\be{c0}
{\bf c_0}  = {\bf c'_0}/gi_0 =  (1,q_0,u_0,v_0)^T =(1,a,b,c)^T.
\ee  
 $a,b$ and $c$ are thus known  from the measured counts in each demodulated state $(q_0,u_0,v_0)$ divided by the continuum intensity.  
For the case studied by SAVV93 (${\bf S= S_1}$, no linear polarization):
\vskip 8pt
\( {\bf c_1}=
\begin{bmatrix}
i_1\\
q_1\\
u_1\\
v_1\\
\end{bmatrix}
\)
\( = 
\begin{bmatrix}
 I_1 + cV_1\\
aI_1 + eV_1\\
bI_1 + fV_1\\
cI_1 + V_1\\
\end{bmatrix}
\)
\vskip 8pt
\noindent where the $(i_1,q_1,u_1,v_1)$ are again all 
relative to $i_0$, the continuum intensity.
These four equations have four unknowns $I_1,V_1,e,f$ which can be solved for algebraically.  This completes the essence of the analysis of SAVV93.  

For the new case ({\bf S=S_2}, negligible circular polarization) we have the measurements 
\vskip 8pt
\( {\bf c_2}=
\begin{bmatrix}
i_2\\
q_2\\
u_2\\
v_2\\
\end{bmatrix}
\)
\( = 
\begin{bmatrix}
 I_2 + aQ_2 + bU_2 \\
 aI_2 + Q_2 + dU_2 \\
 bI_2 - dQ_2 + U_2 \\
 cI_2 -e Q_2 -f U_2 \\
\end{bmatrix}
\)
\vskip 8pt
\noindent which is another set of four equations for the last four
remaining unknowns $I_2, Q_2, U_2$ and $d$.  Although, unlike the
previous cases, these four equations are non-linear in the
unknowns. The equations for $q_2$ and $u_2$ can be used to eliminate
$d$, yielding 3 equations for $I_2,Q_2$ and $U_2$ for example.  But
the equations are quadratic and the closed solutions are very lengthy,
since the elimination of unknown $d$ gives
\be{e}
Q_2^2 + U_2^2 +I_2(aQ_2+bU_2) - u_2U_2 - q_2Q_2 = 0,
\ee
and the remaining equations are of the form $Q_2= \alpha + \beta I_2$, 
$U_2= \gamma+ \delta I_2$.  

This completes the simple formalism proposed here for using specific
atomic transitions to enable accurate and straightforward polarimetry
through a polarizing optical system.

There are several potential difficulties with this proposal that will be discussed in
a later publication.  For now, we address some difficulties in the next section.

\tabletwo

\section{Outlook}

We have reviewed how lines with no linear and/or very small circular
polarization might help us measure polarized light reliably through a
large polarizing telescope/instrument system.  We have adopted the LS
coupling scheme with single, unmixed configurations in this overview,
with the necessary exception of the spin-forbidden transitions leading
to transitions with large $Q,U$ but small or negligible $V$ (section
\pref{subsec:nocirc}).  It remains to be seen 
if lines can be found which are strong enough (big $gf$) but with small
circular polarization to permit the application of the ideas 
presented in section~\pref{subsec:vecpol}.  Further, the 
analysis presented there must be shown to yield well-posed 
physical solutions to equation~(\pref{eq:e}).  
These points will be addressed
in a publication that is currently under preparation.
   
Existing transition data for the ``$V=0$'' spin-forbidden transitions appear to be, for the important
spectrum of \ion{Fe}{1}, entirely from semi-empirical work by
\citet{Kurucz1988}.  It would seem important to revisit {\em ab-initio}
calculations of this ion.  It is possible
that existing data are of insufficient accuracy to provide important
information for application to the kind of solar spectropolarimetry
advocated here.  

The most obvious needs for new atomic data include the following:
\begin{itemize}
\item Infrared lines.  The DKIST will at first light (2019) be equipped with powerful
  spectropolarimeters operating out to 5$\mu$m, yet most reliable
  atomic parameters for the most useful spectral lines in this range
  (wavelengths, mixing coefficients, Land\'e g-factors, oscillator
  strengths) are either  unavailable or
  semi-empirical in nature, the most reliable being measured at  shorter wavelengths.
\item Spin-forbidden lines of iron group neutrals and singly-charged ions.  Further experiments and {\em ab-initio} systematic studies would be especially useful to
obtain reliable atomic parameters for the lines matching the conditions given by equation~(\pref{eq:nov}). Focus might be placed upon both magnetically-sensitive IR lines, and near-UV lines which might achieve 
the highest angular resolutions.
\end{itemize}

Conversely, it is likely that solar physics will provide new
constraints on atomic calculations as the polarization characteristics
of many lines are measured for the first time in infrared regions.
Thus, mixing coefficients (equations~\pref{eq:mixing} and
\pref{eq:mixingg}) can be assessed through measurements of
polarization of atomic transitions at high sensitivity, for comparison
with correlation calculations in complex systems.

%

%
%
In Table 2 we list pairs of atomic transitions of abundant ions in
which both lines of type $``Q,U=0"$ as well as lines of type $``V=0"$
could be observed over a limited spectral range.  We call these
``calibration pairs''.  We chose a 2 nm width in constructing this
table in order to list a potentially useful line pair, noting that
this exceeds the spectral range of the current ViSP design for DKIST.  Even so,
it is clear that a small but non-zero number of pairs are available for study.  
It should also be noted that problems of both
blending and weak magnetic sensitivity are improved by moving to IR
wavelengths, for which several ``$V=0$'' spin-forbidden 
transitions of \ion{Fe}{1} have 
been computed by Kurucz.  Perhaps further calculations and 
laboratory work up
to 5 $\mu$m is warranted.  

Lastly, generally speaking the two lines of each calibration pair are
formed in different regions of the Sun's atmosphere, with perhaps some
overlap.  Therefore, the particular measurements of ${\bf S_i}$
determined above must be augmented with other data to determine the
vector magnetic field from a particular region.  But the main point here
is that the six coefficients $a$-$f$ of {\bf X} are determined, and
can be applied to any line close enough in wavelength to each
``calibration" pair.

\acknowledgments

The author is very grateful to the ASOS committee for providing funds to be able to attend the 12th ASOS meeting in S\~ao Paolo.  The manuscript was greatly improved through helpful discussions and notes from R. Casini, T. del Pino Aleman, and A. Sainz-Dalda.  The author thanks V. Martinez-Pillet and J. Sanchez-Almeida for helpful conversations.  
The data shown in Figures 1 and 2 were acquired with the help of C. Beck and the NSO observers at the Dunn Solar Telescope, operated by the National Solar Observatory, at Sacramento Peak Observatory in Sunspot, NM.


\begin{thebibliography}{0}%
\makeatletter
\providecommand \@ifxundefined [1]{%
 \@ifx{#1\undefined}
}%
\providecommand \@ifnum [1]{%
 \ifnum #1\expandafter \@firstoftwo
 \else \expandafter \@secondoftwo
 \fi
}%
\providecommand \@ifx [1]{%
 \ifx #1\expandafter \@firstoftwo
 \else \expandafter \@secondoftwo
 \fi
}%
\providecommand \natexlab [1]{#1}%
\providecommand \enquote  [1]{``#1''}%
\providecommand \bibnamefont  [1]{#1}%
\providecommand \bibfnamefont [1]{#1}%
\providecommand \citenamefont [1]{#1}%
\providecommand \href@noop [0]{\@secondoftwo}%
\providecommand \href [0]{\begingroup \@sanitize@url \@href}%
\providecommand \@href[1]{\@@startlink{#1}\@@href}%
\providecommand \@@href[1]{\endgroup#1\@@endlink}%
\providecommand \@sanitize@url [0]{\catcode `\\12\catcode `\$12\catcode
  `\&12\catcode `\#12\catcode `\^12\catcode `\_12\catcode `\%12\relax}%
\providecommand \@@startlink[1]{}%
\providecommand \@@endlink[0]{}%
\providecommand \url  [0]{\begingroup\@sanitize@url \@url }%
\providecommand \@url [1]{\endgroup\@href {#1}{\urlprefix }}%
\providecommand \urlprefix  [0]{URL }%
\providecommand \Eprint [0]{\href }%
\providecommand \doibase [0]{http://dx.doi.org/}%
\providecommand \selectlanguage [0]{\@gobble}%
\providecommand \bibinfo  [0]{\@secondoftwo}%
\providecommand \bibfield  [0]{\@secondoftwo}%
\providecommand \translation [1]{[#1]}%
\providecommand \BibitemOpen [0]{}%
\providecommand \bibitemStop [0]{}%
\providecommand \bibitemNoStop [0]{.\EOS\space}%
\providecommand \EOS [0]{\spacefactor3000\relax}%
\providecommand \BibitemShut  [1]{\csname bibitem#1\endcsname}%
\let\auto@bib@innerbib\@empty
\end{thebibliography}%


\begin{thebibliography}{}

\bibitem[\protect\astroncite{{Babcock}}{1953}]{Babcock1953}
{Babcock}, H.~W.: 1953,
\newblock {\em Astrophys.\ J.\/} {\bf 118}, 387

\bibitem[\protect\astroncite{{Babcock} and
  {Babcock}}{1952}]{Babcock+Babcock1952}
{Babcock}, H.~W. and {Babcock}, H.~D.: 1952,
\newblock {\em Publ.\ Astron.\ Soc.\ Pac.\/} {\bf 64}, 282

\bibitem[\protect\astroncite{{Berger} {\em et~al.}}{1998}]{Berger+others1998}
{Berger}, T.~E., {L{\"o}fdahl}, M.~G., {Shine}, R.~A., and {Title}, A.~M.:
  1998,
\newblock {\em Astrophys.\ J.\/} {\bf 506}, 439

\bibitem[\protect\astroncite{{Cameron} {\em
  et~al.}}{2011}]{Cameron+Voegler+Schuessler2011}
{Cameron}, R., {V{\"o}gler}, A., and {Sch{\"u}ssler}, M.: 2011,
\newblock {\em \aap\/} {\bf 533}, A86

\bibitem[\protect\astroncite{{Casini} and {Landi
  Degl'\-Innocenti}}{2008}]{Casini+Landi2008}
{Casini}, R. and {Landi Degl'\-Innocenti}, E.: 2008,
\newblock {\em Plasma Polarization Spectroscopy\/}, Chapt. 12. Astrophysical
  Plasmas,  247,
\newblock Springer

\bibitem[\protect\astroncite{Cowan}{1981}]{Cowan1981}
Cowan, R.~D.: 1981,
\newblock {\em The Theory of Atomic Structure and Spectra\/},
\newblock University of California Press, Berkeley CA

\bibitem[\protect\astroncite{{Denker} {\em et~al.}}{2012}]{Denker+others2012}
{Denker}, C., {Lagg}, A., {Puschmann}, K.~G., {Schmidt}, D., {Schmidt}, W.,
  {Sobotka}, M., {Soltau}, D., {Strassmeier}, K.~G., {Volkmer}, R., {von der
  Luehe}, O., {Solanki}, S.~K., {Balthasar}, H., {Bello Gonzalez}, N.,
  {Berkefeld}, T., {Collados Vera}, M., {Hofmann}, A., and {Kneer}, F.: 2012,
\newblock {\em IAU Special Session\/} {\bf 6}, E2.03

\bibitem[\protect\astroncite{Dufton and Kingston}{1981}]{Dufton+Kingston1981}
Dufton, P.~L. and Kingston, A.~E.: 1981,
\newblock {\em Adv. At. Molec. Phys.\/} {\bf 17}, 355

\bibitem[\protect\astroncite{Gabriel and Jordan}{1971}]{Gabriel+Jordan1971}
Gabriel, A.~H. and Jordan, C.: 1971,
\newblock {\em Case Studies in Atomic Collision Physics\/}, Chapt.~4,
  210--291,
\newblock North-Holland

\bibitem[\protect\astroncite{{Goode} and {Cao}}{2012}]{Goode+Cao2012}
{Goode}, P.~R. and {Cao}, W.: 2012,
\newblock in T.~R. {Rimmele}, A. {Tritschler}, F. {W{\"o}ger}, M. {Collados
  Vera}, H. {Socas-Navarro}, R. {Schlichenmaier}, M. {Carlsson}, T. {Berger},
  A. {Cadavid}, P.~R. {Gilbert}, P.~R. {Goode}, and M. {Kn{\"o}lker} (Eds.),
  {\em Second ATST-EAST Meeting: Magnetic Fields from the Photosphere to the
  Corona.\/}, Vol. 463 of {\em Astronomical Society of the Pacific Conference
  Series\/},  357

\bibitem[\protect\astroncite{Harrington and
  Sueoka}{2016}]{Harrington+Sueoka2016}
Harrington, D. and Sueoka, S.~R.: 2016,
\newblock Vol. 9912, Proc SPIE

\bibitem[\protect\astroncite{Keil {\em et~al.}}{2009}]{Keil+Rimmele+Wagner2009}
Keil, S., Rimmele, T., and Wagner, J.: 2009,
\newblock {\em Earth, Moon and Planets\/} {\bf 104}, 77

\bibitem[\protect\astroncite{Kurucz}{}]{Kurucz1988}
Kurucz, R.~L.,
\newblock in M. McNally (Ed.), {\em Trans. IAU, XXB\/}, Dordrecht: Kluwer,
  p.~168

\bibitem[\protect\astroncite{{Landi Degl'Innocenti}}{2013}]{Landi2013}
{Landi Degl'Innocenti}, E.: 2013,
\newblock {\em Memorie della Societa Astronomica Italiana\/} {\bf 84}, 391

\bibitem[\protect\astroncite{{Landi Degl'Innocenti} and
  {Landolfi}}{2004}]{Landi+Landolfi2004}
{Landi Degl'Innocenti}, E. and {Landolfi}, M.: 2004,
\newblock {\em {Polarization in Spectral Lines}\/}, Vol. 307 of {\em
  Astrophysics and Space Science Library\/}

\bibitem[\protect\astroncite{{Landstreet}}{1969}]{Landstreet1969}
{Landstreet}, J.~D.: 1969,
\newblock {\em Publ.\ Astron.\ Soc.\ Pac.\/} {\bf 81}, 896

\bibitem[\protect\astroncite{{Lites}}{1993}]{Lites1993}
{Lites}, B.~W.: 1993,
\newblock {\em Solar Phys.\/} {\bf 143}, 229

\bibitem[\protect\astroncite{{Makita}}{1986}]{Makita1986}
{Makita}, M.: 1986,
\newblock {\em Solar Phys.\/} {\bf 106}, 269

\bibitem[\protect\astroncite{Parker}{1979}]{Parker1979}
Parker, E.~N.: 1979,
\newblock {\em Cosmical magnetic fields\/},
\newblock Clarendon Press, Oxford


\bibitem[\protect\astroncite{{Parker}}{2009}]{Parker2009}
{Parker}, E.~N.: 2009,
\newblock {\em \ssr\/} {\bf 144}, 15

\bibitem[\protect\astroncite{Rimmele and Marino}{2011}]{Rimmele+Marino2011}
Rimmele, T. and Marino, J.: 2011,
\newblock {\em LRSP\/} {\bf 8}, 2

\bibitem[\protect\astroncite{{Sanchez Almeida} and {Vela
  Villahoz}}{1993}]{Sanchez-Almeida+Vela-Villahoz1993}
{Sanchez Almeida}, J. and {Vela Villahoz}, E.: 1993,
\newblock {\em Astron.\ Astrophys.\/} {\bf 280}, 688

\bibitem[\protect\astroncite{Seagraves and Elmore}{1994}]{Seagraves+Elmore1994}
Seagraves, P.~H. and Elmore, D.~F.: 1994,
\newblock {\em Proc. SPIE\/} {\bf 2265}, 231

\bibitem[\protect\astroncite{{Vela Villahoz} {\em
  et~al.}}{1994}]{Vela-Villahoz+others1994}
{Vela Villahoz}, E., {Sanchez Almeida}, J., and {Wittmann}, A.~D.: 1994,
\newblock {\em Astron.\ Astrophys.\/} 103

\end{thebibliography}
\end{document}